\documentclass{elsart}
\usepackage{amsfonts,amsmath,amssymb,array,graphicx,tabularx}
\usepackage[mathscr]{eucal}

\newcommand{\im}{\mathop{\mathrm{Im}}}

\begin{document}

\begin{frontmatter}

\title{An adiabatic mode time-dependent parabolic equation}

\author{M.Yu.~Trofimov}\ead{trofimov@poi.dvo.ru}

\address{Il'ichev Pacific oceanological institute, Baltiyskay St. 43, Vladivostok, 41, 690041, Russia }
\begin{abstract}
An adiabatic mode time-dependent parabolic equation for the 3D problems of underwater acoustic
propagation problem is derived by the multiple-scale method. The Cauchy problem for this equation is considered.
\end{abstract}

\end{frontmatter}

\section{Introduction}
\par
An adiabatic mode continuous wave parabolic
equation was derived by the factorization of
horizontal Helmholtz operator in the recent paper of
M. Collins \cite{coll}.
Another form of this equation was obtained by the generalized multiple-scale method
in our paper \cite{trof1}. The use of multiple-scale method is especially important
in nonstationary problems because it automatically gives the correct variables for the
amplitude and phase of the acoustic field \cite{trof2}.
\par
In this paper we derive by the multiple-scale method an adiabatic mode time-dependent
parabolic equation and the characteristic equations for the associated Hamilton-Jacobi
equation. We use the characteristic variables to give a reduction of the Cauchy
problem for the derived equation to a family of those for (formally) continuous
wave parabolic equation.

\section{Formulation and derivation}
\par
We start with the acoustic wave equation for nonstationary media
\begin{equation} \label{1}
\mbox{div}\,\left(\frac{1}{\rho}\,\mbox{grad}\, p \right) -
\frac{\partial}{\partial t}\left(\frac{1}{\rho} n^2
\frac{\partial p}{\partial t}\right) = 0\,,
\end{equation}
where $p$ is the acoustic pressure, $\rho$ is the density,
$n = 1/c$ is the index of refraction, $c$ is the sound speed.  The
variables are nondimensional, based on a length
scale $\bar l$ , a time scale $\bar l/\bar c$  (where $\bar c$  is a typical
value of sound speed), and a density scale $\bar \rho$  (a
typical value of the density). Equation (\ref{1}) is
considered in a domain
$-H \le z \le h(x,y,t)$ ,
where $H$ is the depth of a fictitious boundary in the bottom bed,
with the boundary conditions
\begin{equation} \label{2}
p = 0 \quad \mbox{at} \quad z = h(x,y,t), \qquad
\frac{\partial p}{\partial z} = 0 \quad \mbox{at} \quad z = -H\,.
\end{equation}
\par
We suppose that the density and the
index  of refraction are piece-wise continuous
with respect to $z$ with discontinuities at surfaces
$z=h^{(l)}(x,y,t)$, $l=1,\ldots,N$. The typical examples of such surfaces
are the ocean bottom and the seismic boundaries in it.
At these surfaces we set the interface boundary condition
\begin{equation} \label{2a}
p_{+} = p_{-}\,,\qquad \left(\frac{1}{\rho}\frac{\partial p}{\partial\nu}\right)_{+} =
\left(\frac{1}{\rho}\frac{\partial p}{\partial\nu}\right)_{-}\,,
\end{equation}
where $\partial/\partial\nu$  denotes the normal derivative. By the subscripts
``$+$'' and ``$-$'' are marked the limits of variables  from above and below the surface.
\par
    Let $\epsilon$  be a small parameter. We introduce
the slow variables $T=\epsilon t$, $X = \epsilon x$, $Y = \epsilon^{1/2}y$,
the fast variable $\eta=(1/\epsilon)\theta(X,Y,z,T)$, and postulate the expansions
\begin{eqnarray*}
\rho & = & \rho_0(X,T,z) + \epsilon\rho_1(X,Y,T,z)\,,\\
n & = & n_0(X,T,z) + \epsilon n_1(X,Y,T,z)\,,\\
h & = & \epsilon h_1(X,Y,T)\,,\\
h^{(l)} & = & h^{(l)}_0(X,T) + \epsilon h^{(l)}_1(X,Y,T)\,,\quad l = 1,\ldots,N\,,\\
p & = & p_0(X,Y,T,z,\eta) + \epsilon p_1(X,Y,T,z,\eta)+\ldots\,.
\end{eqnarray*}
Substituting these expansions to the (\ref{1}), (\ref{2}) and (\ref{2a})
and replacing partial derivatives by the rule
$$
\frac{\partial }{\partial X} \rightarrow \frac{\partial }{\partial X}
+ \frac{1}{\epsilon}\theta_X\frac{\partial }{\partial \eta}\,,
$$
and analogously for the variables $z$, $Y$ and $T$, we
obtain by collecting terms of the same order of $\epsilon$   a
sequence of boundary value problems.
\par
    The problems at $O(\epsilon^{-2})$ and $O(\epsilon^{-1})$ show
that  $\theta$ is independent from $z$ and $Y$.
\par
 At $O(\epsilon^{0})$  we obtain
\begin{equation*} 
\frac{1}{\rho_0}\left(\theta_X\right)^2 p_{0\eta\eta} +
\left(\frac{1}{\rho_0} p_{0z}\right)_z - \frac{1}{\rho_0} n_0^2
\left(\theta_T\right)^2 p_{0\eta\eta} =0\,.
\end{equation*}
The separation of variables $\eta$   and $z$ leads, up to
the gauge transformation of the phase $\theta$ , to a
solution of the form
\begin{equation*}
p_0 = A(X,Y,T)\phi(z,X,T)e^{i\eta}.
\end{equation*}
\par
   The function $\phi$   is a solution to the spectral
boundary value problem
\begin{equation}\label{5}
 \left(\frac{1}{\rho_0} \phi_z\right)_z + \frac{1}{\rho_0} n_0^2
\omega^2 \phi =
\frac{1}{\rho_0} k^2 \phi
\end{equation}
with the boundary conditions
$$
\phi = 0 \quad\text{at}\quad z=0
$$
$$
\phi_z = 0\quad\text{at}\quad z = -H
$$
and the interface conditions
$$\phi_{+} = \phi_{-}\,,$$
$$\left(\rho_0^{-1}\partial \phi/\partial z\right)_{+} =\left(\rho_0^{-1}\partial \phi/\partial z\right)_{-}
\quad\text{at} \quad z=h^{(l)}_0\,,\quad l=1,\ldots, N\,.$$
 Here $k = \theta_X$    is the local wave number and $\omega = -\theta_T$  is the
local frequency, the spectral parameter of (\ref{5}) is identified with $k^2$.
The problem (\ref{5}) can be considered as the Hamilton-Jacobi equation:
\begin{equation} \label{7}
\left(\theta_X\right)^2 = {\mathcal F}^2\left(\theta_T,X,T\right)\,.
\end{equation}
The phase $\theta$  can be complex with $\im\theta\ge 0$ (the dissipation condition), so
for the normalization and orthogonality conditions we use
the non-Hermitian inner product
\begin{equation} \label{8}
(\phi,\psi) = \int_{-H}^0\frac{1}{\rho_0}\phi\psi\,dz\,.
\end{equation}
which accords with the biorthogonality conditions \cite{codd-lev}
\par
The consideration of the solvability condition of the boundary value problem at $O(\epsilon)$,
formulated as the orthogonality condition with respect to  (\ref{8}),
 gives the parabolic equation
\begin{equation} \label{10}
A_X + \frac{1}{c_g} A_T -i\frac{1}{2 k} A_{YY} + \alpha A = 0\,,
\end{equation}
where
\begin{equation*} 
\begin{split}
& \alpha = \frac{1}{2}\left(\ln k\right)_X + \frac{1}{2}\left(\ln
k\right)_T\cdot\frac{1}{c_g} +
\frac{1}{2}\left(\frac{1}{c_g}\right)_T \\ & \qquad -
i\frac{1}{2k}\int_{-H}^0
\frac{\rho_1}{\rho_0^2}\left[\left(\phi_z\right)^2 + k^2\phi^2 -
\omega^2 n_0^2\phi^2\right]\,dz \\ & \qquad -
i\frac{\omega^2}{k}\int_{-H}^0\frac{n_1}{\rho_0}n_0\phi^2\,dz -
\left.\frac{i}{2k\rho_0}h_1\cdot\left(\phi_z\right)^2\right|_{z=0}
\\ &
\qquad + \frac{i}{2k}\sum^N_{l=1}\left.\left(\frac{1}{\rho_0}\phi_z\right)^2_{+}\cdot
h_1^{(l)}\cdot
\left(\rho_{0+}-\rho_{0-}\right)\right|_{z=h_0^{(l)}}  \\ &
\qquad - \frac{i}{2k}\sum^N_{l=1}h_1^{(l)}\phi^2\left[ k^2
\left(\frac{1}{\rho_{0+}}-\frac{1}{\rho_{0-}}\right) -\right. \\ &
\qquad\qquad\qquad\qquad\qquad\qquad\omega^2\left.\left.
\left(\frac{n_{0+}^2}{\rho_{0+}}-\frac{n_{0-}^2}{\rho_{0-}}\right)\right]
\right|_{z=h_0^{(l)}}\,,
\end{split}
\end{equation*}
$c_g$  is the group velocity of the  normal mode under consideration.
\par
    For the waves propagating to the positive
$X$-direction the Hamilton-Jacobi equation (\ref{7}) takes the
form
\begin{equation} \label{12}
\theta_X = {\mathcal F}\left(\theta_T,X,T\right)\,.
\end{equation}

\section{The Cauchy problem}
\par
For equations (\ref{12}), (\ref{10}) we consider the Cauchy
problem with the initial  conditions $\theta(X=0,T)=\theta_0(T)$, $A(X=0,Y,T) = A_0(Y,T)$.
For equation (\ref{12}) the solution of this problem is
given by the method of characteristics
\begin{equation*}
\theta(X,T(X)) = \theta_0(\tau) + \int_0^X\left(k-\frac{\omega}{c_g}\right)\,ds\,,
\end{equation*}
where $(T(s),\omega(s))$  is a solution of the characteristic system
\begin{equation*}
\begin{split}
& \frac{d\omega}{ds} = -{\mathcal F}_T\,,\\
& \frac{dT}{ds} = {\mathcal F}_\omega = \frac{1}{c_g}\,,
\end{split}
\end{equation*}
with initial conditions $T(0) = \tau$, $\omega(0) = -\theta_T(0)$.
The expression for the group velocity is well
known:
$$
c_g = \left(\frac{\omega}{k}\int_{-H}^0\frac{n_0^2}{\rho_0}\phi^2\,dz\right)^{-1}\,.
$$
The expression for  ${\mathcal F}_T$ is given by the formula
\begin{equation*}
\begin{split}
& {\mathcal F}_T = \frac{1}{2k}\int_{-H}^0\frac{1}{\rho_0^2}\phi^2
\left[\left(\omega^2 n_0^2\right)_T\rho_0 + k^2\rho_{0T} - \omega^2
n_0^2\rho_{0T}\right]
\,dz \\
& \qquad + \frac{1}{2k}\int_{-H}^0\frac{\rho_{0T}}{\rho_0^2}\left(\phi_z\right)^2\,dz \\
& \qquad - \frac{1}{2k}\sum^N_{l=1}\left.\left(\frac{1}{\rho_0}\phi_z\right)^2_{+}
\cdot h_{0T}^{(l)}\cdot
\left(\rho_{0+}-\rho_{0-}\right)\right|_{z=h_0^{(l)}}  \\
& \qquad + \frac{1}{2k}\sum^N_{l=1}h_{0T}^{(l)}\phi^2\left[ k^2
\left(\frac{1}{\rho_{0+}}-\frac{1}{\rho_{0-}}\right) \right. \\
& \qquad\qquad\qquad\qquad\qquad\qquad - \omega^2\left.\left.
\left(\frac{n_{0+}^2}{\rho_{0+}}-
\frac{n_{0-}^2}{\rho_{0-}}\right)\right]\right|_{z=h_0^{(l)}}\,.
\end{split}
\end{equation*}
\par
The Cauchy problem for equation (\ref{10}) can be
solved by passing to the characteristic variables $s$, $\tau$. In these variables  equation (\ref{10})
has a more simple form
\begin{equation*}
A_s - i\frac{1}{2k}A_{YY} + \alpha A = 0\,.
\end{equation*}
The solution in initial variables is recovered then as
$A(X,T) = A(s,\tau)$  at $s=X$, where $T = T(X)$
is a solution of the characteristic system
with the  initial condition $T(0)=\tau$. Here
$\tau$  is simply a parameter, so we have formally a
family of Cauchy problems for the continuous wave form of
adiabatic mode parabolic equation (the expression for the coefficient is, of course, different), which
was considered in details in \cite{trof1}.

\section{Conclusion}
In this paper the derivation of  the new adiabatic mode parabolic equation (\ref{10}) is
outlined and the explicit expressions for its coefficients are given.
The Cauchy problem for this equation is proposed to solve by passing to the characteristic variables
and for the corresponding characteristic system the explicit expressions for its coefficients are given.

\end{document}